# PRAXIS-VirtualCell: A Programmable and Trustworthy Framework for Agentic Virtual Cell Experiments

Zhenyu Ma[1], Xukai Jiang[1]*

[1]National Glycoengineering Research Center, Shandong University, Qingdao 266237, China.

***Correspondence**

Xukai Jiang, Email: xukai.jiang@sdu.edu.cn

ABSTRACT:

Virtual cells are evolving from single-task predictive models toward programmable biological simulation systems, yet heterogeneous data, models, and validation evidence still lack a unified organizational framework. Here, we present PRAXIS-VirtualCell, a modular framework that organizes biological data, predictive models, perturbations, adapters, execution environments, and validation evidence to enable reproducible and auditable virtual experiments. The system supports cross-species tasks spanning *Escherichia coli*, *Saccharomyces cerevisiae*, and human K562 cells, while using biological contracts and evidence-aware execution to distinguish supported predictions from extrapolation and abstention. By further integrating agentic orchestration, PRAXIS-VirtualCell automatically translates natural-language questions into traceable virtual experiments, providing a unified runtime foundation for trustworthy and scalable Virtual Cell systems.

## Introduction

Building a "virtual cell" capable of reproducing cellular states in silico, responding to external perturbations, and predicting phenotypic changes has long been a major goal of computational biology.[1] Early whole-cell modeling approaches primarily relied on explicit mechanistic descriptions of biological processes such as metabolism, transcription, translation, DNA replication, and the cell cycle, coupling multiple mathematical models to simulate cellular behavior as an integrated system. For example, the *Mycoplasma genitalium* whole-cell model developed by Karr et al. integrated multiple cellular processes within a unified simulation framework and demonstrated the feasibility of predicting cellular phenotypes from genotype.[1] However, such models typically depend on extensive manually curated mechanistic knowledge and kinetic parameters, and their construction is highly specific to particular organisms and biological systems. This makes them difficult to scale directly to complex eukaryotic cells and to the rapidly expanding space of multi-omics data. In recent years, the rapid accumulation of large-scale single-cell sequencing, Perturb-seq, spatial omics, and high-throughput phenotyping data, together with advances in deep learning and generative artificial intelligence, has begun to shift the Virtual Cell paradigm from explicitly reconstructing all cellular mechanisms toward learning cellular states and perturbation responses directly from data.[2,3,8,14]

This transition has first driven the development of a broad range of artificial intelligence models for cellular state representation and perturbation prediction. Single-cell foundation models such as Geneformer, scGPT, and scFoundation are pretrained on tens of millions of single-cell transcriptomes with the aim of learning generalizable cellular representations that can be transferred to tasks including cell-type annotation, gene regulation, drug response, and genetic perturbation.[21–24] Models such as GEARS go a step further by directly learning mappings from initial cellular states and genetic perturbations to post-perturbation transcriptional states.[5] These approaches have substantially expanded the capabilities of *in silico* cellular perturbation, enabling computational screening of large numbers of candidate interventions before experimental validation.[17–20] However, recent systematic evaluations have also shown that single-cell foundation models do not necessarily outperform simple linear or nearest-neighbor baselines in tasks such as perturbation prediction, and that their advantages are often restricted to particular datasets, task definitions, and evaluation metrics.[6,7] These findings suggest that a substantial gap remains between a high-performing single-task predictor and a Virtual Cell capable of reliably executing open-ended virtual experiments.

Accordingly, the goal of Virtual Cell research is increasingly shifting from building individual predictive models toward constructing computational systems that can represent cellular states, accept interventions, invoke models operating at different biological scales, and generate diverse biological readouts. The recently proposed Virtual Yeast framework decomposes yeast-cell complexity into multiple functional modules, coordinates domain-specific tools using large language models, and establishes a model–experiment feedback loop through active experimentation. Related work on AI-driven digital organisms has further proposed modular and interoperable foundation models for describing multiscale biological systems spanning molecules, cells, and even whole organisms. AIDO Cell, introduced in 2026, further emphasizes a world-model perspective in which internal cellular states are continuously maintained while different interventions and readouts are executed within a shared system. Together, these developments suggest that the next generation of Virtual Cells will likely no longer be synonymous with any single neural network, but will instead take the form of executable scientific systems composed of data, state representations, predictive models, perturbation operators, reasoning systems, and experimental feedback.[2,3,25,26]

However, once a Virtual Cell expands from a single model into a multi-model, multi-dataset, and multi-task system, a new foundational question emerges: how can heterogeneous biological data and computational models be combined into a virtual experiment in a way that is explicit, reproducible, and constrained by biologically meaningful boundaries?[2,3] Different models are typically developed for different species, cell types, experimental conditions, feature spaces, perturbation types, and output definitions.[27–32] Even if two models can both be successfully invoked by a software system, this does not mean that they can answer the same biological question. Likewise, the fact that a dataset can be loaded and a model can complete a forward pass does not by itself establish that the resulting data–model combination is scientifically meaningful. Current evaluations of Virtual Cells remain largely focused on predictive performance, while cross-dataset, cross-condition, and cross-species generalization, together with training-data contamination, task coverage, and result calibration, are becoming increasingly important dimensions of evaluation.[6,7,46] For a truly open-ended Virtual Cell, therefore, the system must answer not only "what is the predicted outcome?" but, before that, "can this question be predicted reliably given the available data, models, and evidence?"

This requirement also makes abstention a scientific outcome that is as important as prediction itself.[45] Real-world virtual experiments will inevitably encounter genes, cellular states, environmental conditions, and perturbation combinations that lie outside the training distribution. When no suitable model is available, when model inputs are incompatible with the experimental state, or when existing benchmarks do not provide sufficient evidence that a model performs reliably under the relevant conditions, forcing the system to return a numerical prediction does not increase the capability of the Virtual Cell; instead, it may obscure the true limits of the model. A trustworthy Virtual Cell therefore needs to explicitly record data provenance, model identity, execution environment, applicable biological conditions, and validation evidence, and to use this information to distinguish supported predictions, out-of-distribution extrapolations, and unsupported queries.[42–44] In other words, the generality of a Virtual Cell should not be measured solely by how many models it can invoke, but also by its ability to recognize the boundaries of its own knowledge and predictive validity.[6,7,45,46]

In this study, we present PRAXIS-VirtualCell, a modular runtime framework for programmable virtual-cell experiments.[2,3] Rather than treating a Virtual Cell as a single general-purpose predictive model, we construct it as an executable system composed of biological data, cell state, perturbation, predictive model, adapter, execution environment, and validation evidence, with explicit biological contracts governing compatibility among datasets, models, and experimental conditions. This framework allows different organisms and tasks to retain their own data representations and specialized models while sharing a common runtime for experiment definition, execution, validation, and result tracking. We further validate this design across heterogeneous prediction tasks involving *Escherichia coli*, *Saccharomyces cerevisiae*, and human K562 cells, and introduce evidence-aware execution to distinguish supported predictions from extrapolations beyond the validated scope and unsupported queries.[4–7] Building on this foundation, agentic orchestration translates natural-language research questions into structured, traceable, and reproducible virtual experiments.[51,52] PRAXIS-VirtualCell thus provides an engineering path from isolated predictive models toward scalable, auditable, and programmable Virtual Cell systems.

## Result

### A modular architecture for programmable virtual cells

To enable Virtual Cells to scale across species, cell types, and prediction tasks, we redefine the Virtual Cell as a modular runtime rather than a single predictive model. Each virtual experiment is decomposed into independently configurable components, including biological data, cell state, perturbation, predictive model, adapter, execution environment, and validation evidence.[2,3] Models therefore function as replaceable modules for specific state transitions or phenotype predictions and can be combined through a shared runtime, transforming a fixed prediction pipeline into a programmable biological experiment system[21–26] (Figure 1a).

Biological data are decoupled from specific predictive models through a structured VirtualCellPack that records species or cell identity, experimental conditions, feature space, perturbation type, controls, omics information, and data provenance.[42–44] Models independently declare their supported species, inputs, perturbations, and outputs.[27–32] This data–model separation allows heterogeneous datasets and specialized models to be integrated without modifying the core runtime and provides an explicit basis for compatibility checking (Figure 1b).

Adapters and execution environments are also managed independently. Adapters translate standardized virtual experiment requests into model-specific inputs and convert outputs back into standardized biological results, while software versions, parameters, and computational environments are recorded separately.[27–32,42–44] This enables executions to be traced to specific combinations of data, model, adapter, and environment and makes individual experiments reproducible rather than merely executable (Figure 1c).

Each Virtual Cell query is represented as an explicit virtual experiment. The runtime defines the biological state and perturbation, checks compatibility among the data, model, and task, executes the selected model when valid, and records the resulting prediction together with its full experimental context. Changes in biological context, model version, or execution conditions therefore produce distinct experiment records rather than context-free predictions (Figure 1d).

Validation evidence is separated from model executability.[6,7] A model that can technically process an input is not necessarily scientifically supported under the corresponding biological condition. PRAXIS-VirtualCell therefore associates models with independent benchmarks, applicability conditions, and validation evidence, allowing the runtime to distinguish supported predictions from unsupported extrapolation and providing the basis for evidence-aware execution and abstention.[42–46]

Together, these components form a modular runtime organized around a unified biological experiment contract.[2,3] New datasets can expand representable cell states, new models can add state transitions or readouts, and new adapters or validation evidence can extend executability and supported scope without rebuilding the system. An upper-layer agent translates biological questions into structured experiment requests, while numerical outputs remain generated by registered scientific models[51,52] (Figure 1e).

The generality of PRAXIS-VirtualCell therefore comes not from a universal predictor, but from a shared runtime that organizes heterogeneous biological states, specialized models, and validation evidence while preserving clear task boundaries and execution identities.[2,3] We next tested whether this runtime could support cross-species Virtual Cell tasks with distinct data representations, perturbation definitions, and output spaces.[4–7]

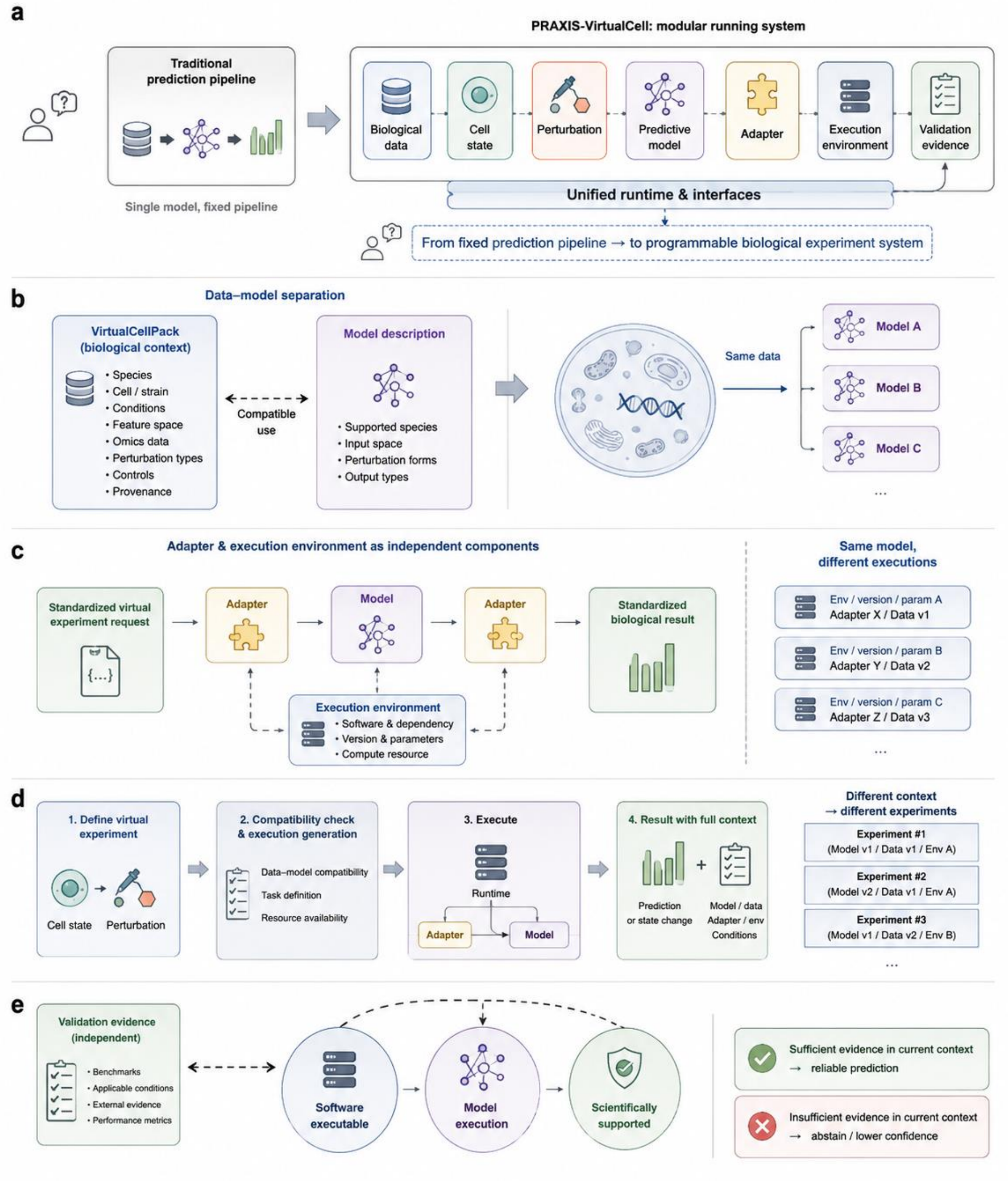

**Figure 1. Modular architecture of PRAXIS-VirtualCell.**

(a) A fixed prediction pipeline is reorganized into a modular Virtual Cell runtime composed of biological data, cell state, perturbation, predictive model, adapter, execution environment, and validation evidence. (b) Biological data and models are described independently and connected through explicit compatibility rules. (c) Adapters and execution environments are managed as independent components, enabling traceable and reproducible model execution. (d) Each query is converted into an explicit virtual experiment, including state definition, compatibility checking, execution, and contextualized result recording. (e) Validation evidence is separated from model executability, allowing the system to distinguish supported predictions from conditions requiring abstention.

**A shared runtime and explicit biological contracts support reproducible Virtual Cell experiments across heterogeneous biological systems**

To test whether the modular architecture could support Virtual Cell tasks with different biological definitions, we implemented three representative systems in PRAXIS-VirtualCell: *Escherichia coli*, *Saccharomyces cerevisiae*, and human K562 cells, covering gene-knockout growth, gene-deletion fitness, and high-dimensional transcriptional responses to genetic perturbation.[8–16] Rather than enforcing a common biological representation, VirtualCellPack records species or cell identity, assay, experimental conditions, feature space, controls, perturbation space, and data provenance, while model contracts define supported inputs, perturbations, and outputs. Task-specific feature alignment and input/output conversion are handled by adapters.[27–32] Thus, the shared runtime standardizes virtual experiment definition, execution, and recording rather than the underlying biological models (Figure 2a).

We first benchmarked two microbial systems against public data. For *E. coli* K-12, iML1515 was matched to LBL FEBA/Keio carbon-source screening data, yielding 26 conditions, 1,339 shared targets, and 34,814 condition–target pairs with feasible wild-type solutions.[34,54,55] All 26 conditions showed positive Spearman correlations, with a median ρ of 0.2237, global ρ of 0.2273, and global Pearson r of 0.6171. Across 1,339 genes, the median target-level Spearman ρ was 0.0403, with 54.1% showing positive correlations. The sucrose condition, which lacked a feasible wild-type solution, was recorded as unsupported rather than forced to produce a prediction.[45,46]

The same workflow was applied to *S. cerevisiae* using iMM904 and public S288C deletion-fitness data.[33,38,56] Across four feasible 48-h conditions, we obtained 722 shared targets and 2,886 condition–target pairs. The condition-level median Spearman $\rho$ was 0.0408, global Spearman $\rho$ was 0.0107, target-level median $\rho$ was 0.0000, and within-condition normalized z-MSE was 1.9380. The very-low-glucose condition was likewise marked as unsupported. Together, these microbial benchmarks show that a single runtime can accommodate distinct species, culture conditions, and readouts while preserving their biological semantics and producing standardized benchmark records.

The K562 task further tested support for multiple predictors and high-dimensional outputs. Under fixed Norman_2019 Golden validation conditions, a simple additive model and GEARS were compared across 30 perturbation conditions and 5,045 output features.[4,5,8–13,53] Their macro MSEs were 0.002231 and 0.004206, respectively, with the additive model performing better in 25/30 conditions. Among 14 single-gene perturbations, the mean MSEs were 0.002072 and 0.002512, with the additive model better in 10/14 conditions. Among 16 combinatorial perturbations, the corresponding values were 0.002369 and 0.005688, with the additive model better in 15/16 conditions. Combinatorial perturbations accounted for approximately 89.6% of GEARS' net excess error, showing that the shared experimental contract preserves perturbation semantics and enables direct condition-level model comparison (Figure 2b).

The runtime also supports independent external evaluation. In GSE220974, the additive model improved Spearman correlation over a global-mean baseline for all five targets, with a mean improvement of +0.04968 and a one-sided exact sign-flip $p = 0.03125$.[14,17–20] In GSE146194, Candidate-2 was evaluated across seven targets and 5,045 features, achieving lower MSE than the comparator for 4/7 targets; the equally weighted mean Candidate-minus-comparator MSE difference was +0.000221. These results further show that predictor performance can vary across biological contexts.

Across the three systems, the benchmark covers 34,814 condition–target pairs in *E. coli*, 2,886 pairs in *S. cerevisiae*, and 30 Golden conditions plus 12 external targets in K562. Because their biological readouts and metrics differ, we do not combine them into a single cross-species ranking.[6,7,14,32] Instead, these results demonstrate that PRAXIS-VirtualCell can organize heterogeneous models, real-data benchmarks, model comparisons, and external validation within a shared runtime while preserving the biological context and execution identity of each virtual experiment.

PRAXIS-VirtualCell further separates validation evidence from execution. Development validation, external stress tests, technical reproducibility, and scientific validation are stored as distinct evidence states.[42–46] Predictions can therefore be marked as supported or unsupported according to the biological contract and validated scope, while deterministic replay is treated as evidence of technical reproducibility rather than biological validity (Figure 2c).

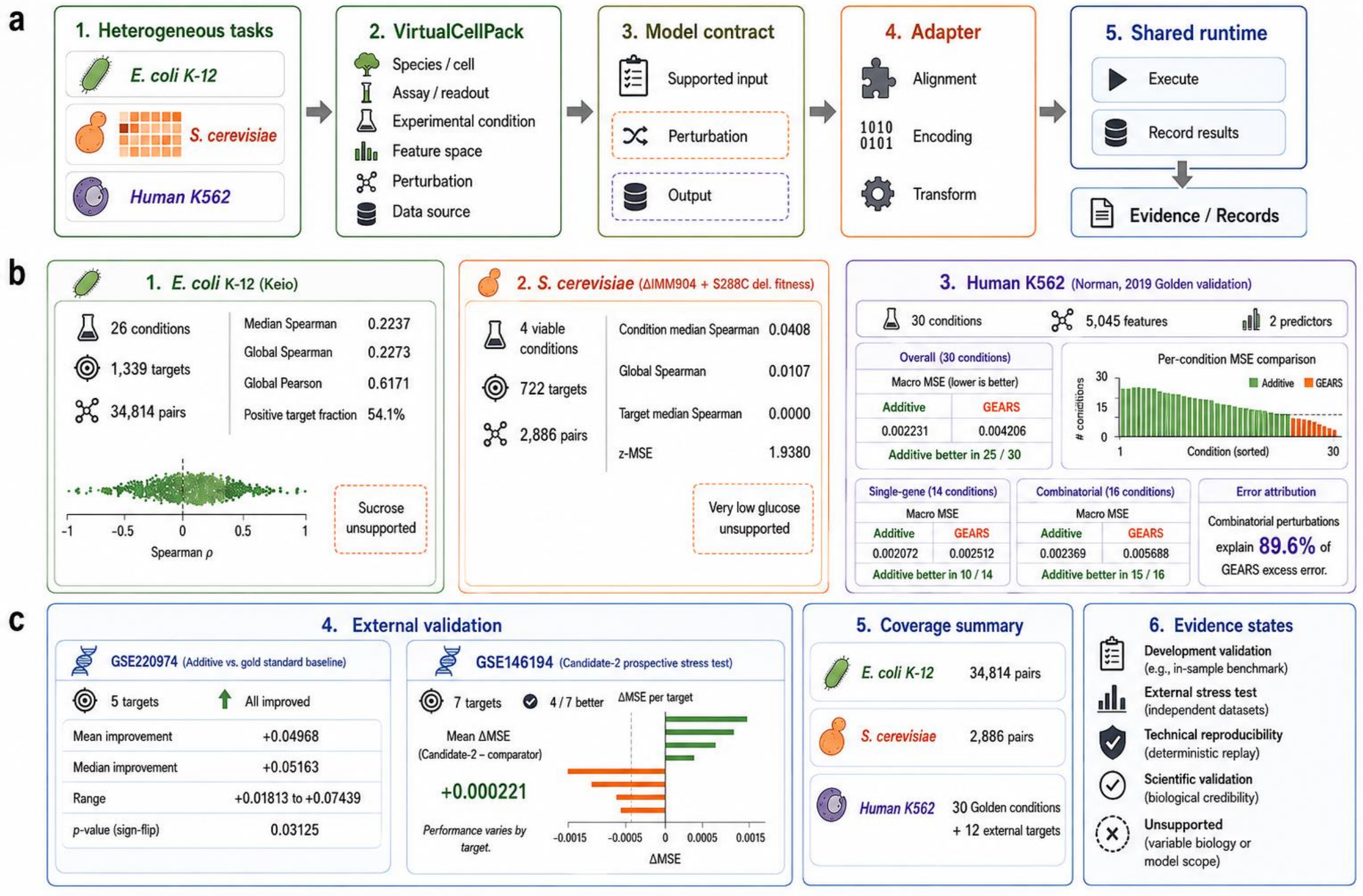

**Figure 2. Shared runtime enables heterogeneous Virtual Cell tasks and unified benchmarking across biological systems.**

(a) Heterogeneous Virtual Cell tasks in *E. coli*, *S. cerevisiae*, and human K562 are organized through VirtualCellPack, model contracts, task-specific adapters, and a shared runtime while preserving task-specific biological representations. (b) The shared runtime supports quantitative benchmarking across *E. coli* knockout growth, *S. cerevisiae* deletion fitness, and K562 genetic perturbation response, including condition-level evaluation, model comparison, and explicit identification of unsupported conditions. (c) Independent external datasets extend validation beyond the development benchmarks, while benchmark coverage and evidence states are recorded separately to distinguish development validation, external stress testing, technical reproducibility, scientific validation, and unsupported predictions.

## Agentic orchestration translates natural-language questions into auditable Virtual Cell experiments

After establishing the modular runtime, biological contracts, and evidence-aware execution, we integrated these components into an agentic Virtual Cell workflow for researcher interaction. Rather than using a large language model to directly predict cellular behavior, the agent parses natural-language questions, routes data and models, constructs virtual experiments, and interprets results[51,52], while numerical prediction and scientific validity remain controlled by the deterministic runtime (Figure 3a).

Each query proceeds through five stages: Ask, Review, Route, Run, and Results.[51,52] A natural-language description of the biological system, perturbation, and target readout is first converted into a structured experiment request. The Review stage exposes the parsed biological context and prediction target for compatibility analysis, after which the Route stage identifies suitable VirtualCellPacks, models, and adapters and generates candidate execution plans (Figure 3b).

The agent and runtime have explicit authority boundaries.[43,44] The language model may assist with parsing, planning, and textual interpretation, but cannot create data, alter feature mappings, generate model outputs, issue validation evidence, or override execution decisions. Compatibility is determined by explicit contracts among the request, VirtualCellPack, model, adapter, and environment, and numerical outputs can only come from registered predictors[6,7]. Supported requests are compiled into deterministic Virtual Experiment DAGs with recorded execution steps and dependencies, keeping the scientific source of truth within structured data and computational components (Figure 3c).

Evidence-aware checks are performed before execution.[45,46] The runtime verifies data availability, model compatibility, execution environment, perturbation scope, and validation evidence. Supported requests return predictions together with their biological context and evidence scope, whereas unmet requirements produce structured abstention rather than a language-generated substitute. Failure states such as Golden-gate failures, external stress-test failures, and scientifically_validated=false are therefore preserved in the final output (Figure 3d).

Each Agent Run produces interconnected records of the report, prediction, module routing, execution identity, evidence, trace, and provenance.[42–44] These records preserve the numerical result or abstention state, selected data and models, execution configuration, validation evidence, and deterministic execution path. The final natural-language report is generated from these structured outputs, allowing researchers to receive not only a prediction, but an auditable virtual experiment record linked to its data, model, conditions, and evidence (Figure 3e).

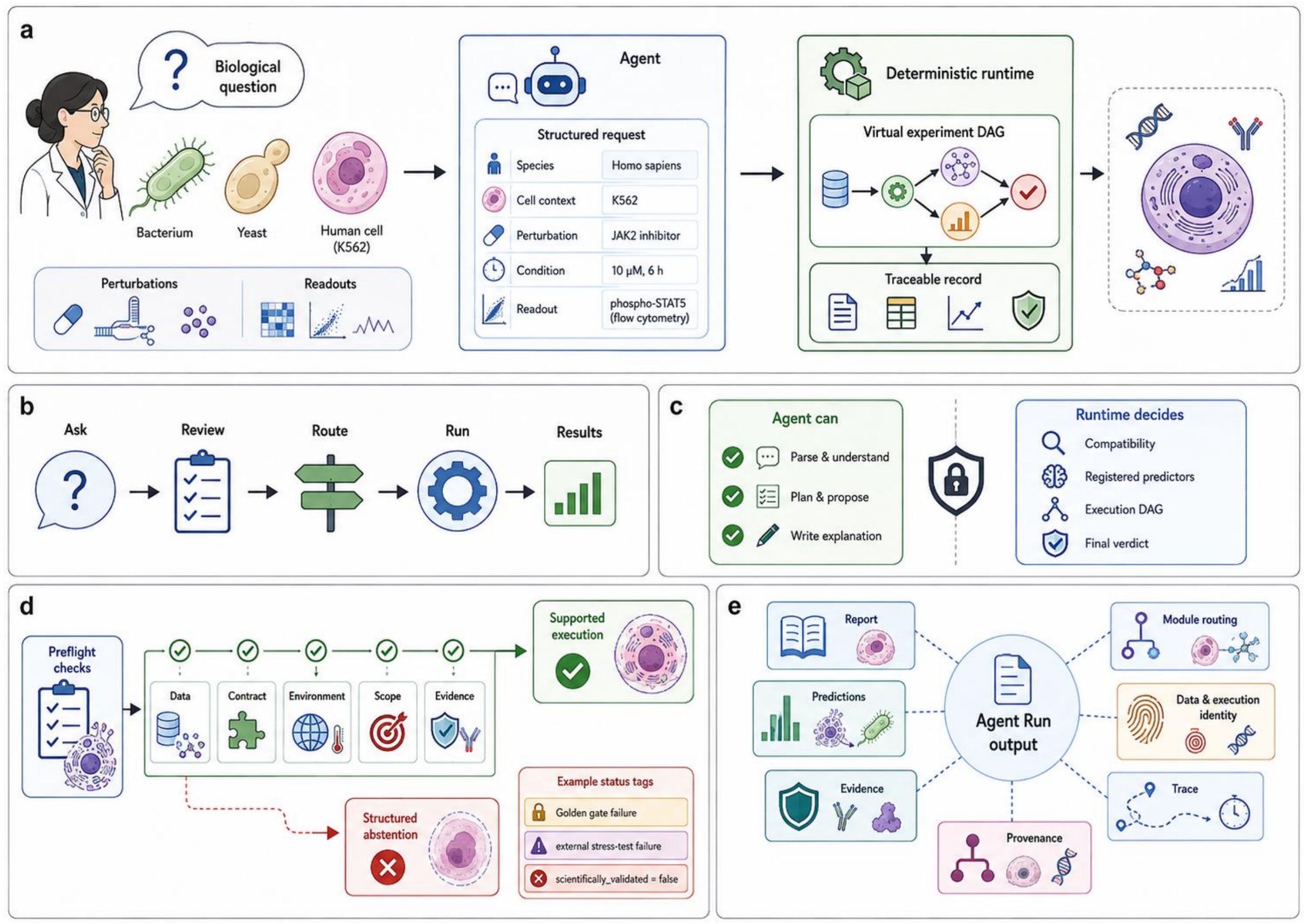

**Figure 5. Agentic workflow and evidence-aware execution in PRAXIS-VirtualCell.**

(a) A natural-language biological question is converted by the agent into a structured virtual experiment request and executed by the deterministic Virtual Cell runtime, which preserves the numerical prediction and scientific decision boundary. (b) Each query proceeds through five explicit stages: Ask, Review, Route, Run, and Results. (c) The agent supports request interpretation, candidate planning, and result explanation, whereas compatibility analysis, registered predictor execution, Virtual Experiment DAG construction, and the final execution verdict remain under runtime control. (d) Evidence-aware preflight checks evaluate data availability, contract compatibility, execution environment, perturbation scope, and validation evidence, leading either to supported execution or structured abstention. (e) Each Agent Run produces a structured virtual experiment record containing the report, predictions, module routing, data and execution identity, evidence, execution trace, and provenance.

## Discussion

In this study, we present PRAXIS-VirtualCell as a programmable Virtual Cell runtime rather than a universal predictive model.[2,3] Biological data, cell state, perturbation, predictive model, adapter, execution environment, and validation evidence are treated as independent components and assembled into executable virtual experiments through a shared biological contract and runtime. In this framework, the generality of a Virtual Cell arises from the ability to continuously integrate, replace, and recombine heterogeneous scientific components.

This design separates model capability from Virtual Cell system capability. Because different species, cell types, and prediction tasks require distinct representations, perturbation spaces, and outputs, PRAXIS-VirtualCell preserves the native representation of individual models while standardizing experiment definition, compatibility checking, execution, and result recording.[27–32] Applications to *E. coli*, *S. cerevisiae*, and K562 demonstrate that heterogeneous Virtual Cell tasks can be supported within the same runtime without changing the core architecture.

A second key principle is the separation of software executability from scientific support.[45,46] Successful computation does not necessarily imply that a prediction is validated for the relevant biological context. Validation evidence is therefore managed independently, allowing the runtime to return either a supported prediction or structured abstention according to the biological contract and available evidence. This enables the system not only to determine how to make a prediction, but also whether a prediction should be made.

The modular design also improves traceability and reproducibility by recording data versions, adapters, feature mappings, software environments, and experimental conditions as part of the execution identity.[42–44] We further introduce agentic orchestration for research-question interpretation, model routing, and result explanation, while compatibility checking, model execution, evidence assessment, and execution verdicts remain within the deterministic runtime.[51,52] The agent therefore translates natural-language questions into structured and auditable virtual experiments rather than directly predicting cellular behavior.

Overall, PRAXIS-VirtualCell extends the Virtual Cell from a predictive model into a programmable biological experimentation system.[2,3] Future foundation models, world models, mechanistic simulations, and multiscale models can be incorporated as modular components, while persistent cell states, uncertainty estimation, and experimental feedback can further expand the system.[21–26,47–50] We envision future Virtual Cells as evolving scientific platforms composed of models, data, states, agents, and validation, rather than as single fixed neural networks.

## Methods

### PRAXIS-VirtualCell architecture and virtual experiment execution

PRAXIS-VirtualCell was implemented as a modular runtime for programmable virtual cell experiments.[2,3] Biological data, cell state, perturbation, predictive model, adapter, execution environment, and validation evidence are managed as independent components and linked through explicit biological contracts. VirtualCellPack describes the biological context of an experiment, including biological identity, assay, conditions, feature and perturbation spaces, controls, and data provenance, whereas the model contract defines the supported biological systems, inputs, perturbations, outputs, required adapters, execution environment, and validation scope.[42–44] Adapters perform identifier mapping, feature alignment, preprocessing, perturbation encoding, and output transformation without extending the biological scope declared by the model.[27–32]

Each request is converted into a structured virtual experiment specifying biological identity, condition, perturbation, and requested readout.[2,3] Before execution, the runtime checks data availability and compatibility across biological identity, feature space, perturbation, readout, adapter, and execution environment, and links the request to available validation evidence.[27–32] Compatible requests are compiled into deterministic virtual experiments and executed by registered predictors.[6,7] Missing data, incompatible models, unsupported perturbations, or unavailable environments are instead recorded as unsupported, abstention, or execution failure. No substitute model or language model is used to generate missing numerical predictions, and software executability alone does not imply scientific validation.[45,46]

Each run records the original request, selected data, model, adapter and environment, experimental conditions, prediction or abstention, evidence state, execution identity, and provenance.[42–44] For deterministic analyses, input ordering, feature order, numerical precision, and runtime configuration were fixed before execution.[46,47] Microbial benchmarks used Python 3.13.12, COBRApy 0.31.1, pandas 2.3.3, and NumPy 2.4.4.[36–38] Linear programming was performed with GLPK through optlang.glpk_interface, with feasibility and integrality tolerances of $10^{-7}$.[35–38] Biomass objectives were maximized using model.slim_optimize() without parsimonious FBA, parameter training, or stochastic sampling. Single-gene microbial simulations were parallelized across four worker processes.

Above this deterministic runtime, PRAXIS-VirtualCell provides a five-stage agentic workflow: Ask, Review, Route, Run, and Results.[51,52] The agent converts natural-language questions into structured requests and can assist with data and model routing and result interpretation, while compatibility decisions, model execution, and evidence states remain controlled by the runtime. The language model cannot modify predictor outputs, invent feature mappings, override compatibility failures, issue validation evidence, or generate substitute predictions for unsupported requests.[43–46] Each Agent Run records the report, prediction or abstention, module routing, execution identity, validation evidence, execution trace, and provenance.

**Microbial quantitative benchmarks**

For the *Escherichia coli* benchmark, we used the iML1515 genome-scale metabolic model, containing 1,516 genes, 2,712 reactions, and 1,877 metabolites, with BIOMASS_Ec_iML1515_core_75p37M as the biomass objective.[34,41] Experimental fitness data were obtained from the LBL FEBA/Keio fitness compendium.[54] Because the experimental strain was *E. coli* BW25113 whereas iML1515 represents MG1655, this benchmark was treated as a cross-strain stress test rather than a strictly strain-matched validation.[39,55]

Carbon-source experiments were selected by retaining conditions with u = TRUE and condition labels containing (C).[54,55] After exact matching to a predefined carbon-source mapping, 27 carbon-source conditions were retained, each represented by two experimental replicates. The original experiments were performed in carbon-free M9 minimal medium at 37 °C under aerobic shaking conditions, with 20 mM of the corresponding carbon source added. Experimental concentrations were not converted directly into flux bounds. Instead, all 27 mapped carbon-source exchange reactions were first closed by setting their lower bounds to 0, after which only the target carbon source was opened with an uptake lower bound of −10.0. Oxygen, ammonia, phosphate, and sulfate exchange bounds were retained from the original SBML model.

Direct string matching between experimental gene identifiers and iML1515 gene IDs yielded 1,339 shared targets. Wild-type biomass was first calculated for each condition, and only conditions with finite positive growth were retained for knockout analysis.[35–38] Single-gene knockouts were then implemented according to the model gene–protein–reaction relationships, followed by re-optimization of the biomass objective.[35–38,55] Predicted fitness was defined as the ratio between knockout and wild-type growth under the same condition,

$$r_{g,c} = \frac{v_{g,c}^{\mathrm{KO}}}{v_c^{\mathrm{WT}}},$$

followed by log transformation,

$$\log_2 r_{g,c}.$$

Negative growth values were truncated to 0 before ratio calculation, and ratios were bounded below by $10^{-9}$ before log transformation. The sucrose condition produced an infeasible wild-type solution and was therefore recorded as unsupported rather than assigned a zero value. The final *E. coli* benchmark contained 26 supported conditions, 1,339 shared targets, and 34,814 condition–target pairs.

For the *Saccharomyces cerevisiae* benchmark, we used the iMM904 genome-scale metabolic model, containing 905 genes, 1,577 reactions, and 1,226 metabolites, with BIOMASS_SC5_notrace as the biomass objective.[33,35–38] Experimental deletion-fitness data were obtained from a published yeast knockout fitness dataset.[56] Five 48-h conditions were considered: 20% glucose, 0.2% glucose, 0.02% glucose, 2% glycerol, and 2% maltose. The corresponding substrate uptake magnitudes in the model were set to 10.0, 0.2, 0.02, 10.0, and 10.0, respectively.

Experimental and model gene identifiers were matched directly.[33,38,56] When multiple observations were available for the same gene–condition pair, the median fitness score was used.[33,38,56] q-values were retained during preprocessing but were not used for additional filtering. The extremely-low-glucose condition produced an infeasible wild-type solution and was recorded as unsupported. The remaining four supported conditions yielded 2,886 condition–target pairs covering 722 unique targets.

Because yeast deletion s-scores do not represent absolute growth rates, model performance was evaluated primarily using rank agreement and within-condition standardized errors rather than direct comparison of absolute values.[33,38] Predictions and observations were independently standardized within each condition using

$$z_{i,c} = \frac{x_{i,c} - \mu_c}{\sigma_c},$$

where $\mu_c$ and $\sigma_c$ denote the mean and population standard deviation for condition $c$, respectively, with $ddof = 0$. Constant vectors were converted to zero, and NaN or infinite values were excluded from correlation analyses. Target-level correlations were calculated only for genes represented in at least three supported conditions.

The current yeast dataset contains observations from S288C, Y55, YPS, and UWOP strains, and the present analysis pipeline does not explicitly apply an S288C-only filter before aggregation. The resulting 2,886-pair benchmark should therefore be interpreted as a multi-strain median benchmark after condition and gene matching rather than a strictly S288C-only analysis.

**K562 genetic perturbation benchmark**

The K562 genetic perturbation benchmark was based on the Norman 2019 CRISPR activation dataset (DOI: 10.7910/DVN/Q2ZV3E).[4] The processed dataset contained 91,205 cells and 5,045 output features, using the EnsemblGene/Ensembl 84 namespace mapped to GRCh38.[8–13] Perturbations were applied for 5 d, with control and perturbation doses defined as 0 and 1, respectively. The original dataset contained 45,989 training, 10,944 validation, and 34,272 test cells, corresponding to 137, 30, and 116 non-control conditions. The Golden benchmark used the 30 non-control conditions from the validation split, including 14 single-gene and 16 combinatorial perturbations.[5,6]

For each perturbation condition $c$, the experimental response was defined as the difference between the mean expression state of perturbed cells and matched control cells from the same split:

$$\Delta x_c = \bar{x}_c - \bar{x}_{\mathrm{control}}.$$

Both model predictions and reference responses were projected onto a fixed order of 5,045 features.[15,32] Feature mapping was locked before model comparison, and models were not allowed to expand the evaluation space through missing-feature imputation or feature reordering.

The additive candidate was implemented as a train-only ridge additive model with an intercept and perturbation-target coefficients.[5,6] For combinatorial perturbations, prediction was obtained by summing the coefficients of the corresponding observed targets, whereas coefficients for targets absent from training were set to zero. The ridge parameter was selected from a predefined grid ranging from 0.01 to 100, with α=0.1\alpha=0.1 selected for the final model. The model was refitted using 104 canonical training conditions covering 68 training targets. It was treated as a development candidate rather than as independently validated scientific evidence. An external additive comparator used α=1.0\alpha=1.0, 137 training conditions, and a 105-target universe; therefore, differences between the two additive models were not interpreted as a simple ridge-parameter ablation.

GEARS was trained strictly on the native training data.[5] Three independent models with random seeds 11, 23, and 37 were trained for 20 epochs with a batch size of 32, learning rate of 0.001, weight decay of 0.0005, hidden size of 64, one GO layer and one gene-GNN layer, decoder hidden dimension of 16, 20 similar genes per graph, a co-expression threshold of 0.4, and a direction loss weight of 0.1; uncertainty prediction was disabled. Native training conditions were further divided into 93 internal-training and 11 internal-validation conditions, and checkpoints were selected according to internal-validation differential-expression MSE. The perturbation vocabulary contained 9,853 genes, while the reconstructed co-expression and GO graphs contained 6,149 and 202,314 edges, respectively. Predictions from the three seeds were averaged in float64 using the fixed 5,045-feature order. Training was performed on Linux x86_64 with Python 3.11.15, PyTorch 2.6.0+cu124, CUDA 12.4, and one NVIDIA A100-SXM4-40GB GPU. Evaluation was performed deterministically on CPU. GEARS produced native predictions for all 30 Golden conditions, corresponding to a coverage of 30/30 with no fallback predictions.

For evaluation, MSE was calculated across all 5,045 features for each Golden condition, and macro MSE was obtained by averaging condition-level MSE values with equal weight.[6,7] Metrics were reported separately for single-gene perturbations, combinatorial perturbations, and all 30 Golden conditions, without weighting by the number of cells in each condition. The Golden benchmark was used as development validation. Successful model inference and final evidence status were recorded independently, such that successful execution did not automatically indicate scientific validation.

The current internal contract retains both cell_identity_id=CVCL_0004 and the upstream field source_cell_type_value=A549. Because these annotations conflict with the stated K562 benchmark identity, this provenance inconsistency is retained explicitly pending confirmation of whether the A549 annotation represents an upstream metadata error or another source-specific field.

**Independent external evaluation and statistical analysis**

Two public datasets independent of the Norman development benchmark were used for external stress testing of the K562 models.[14] GSE220974 is a K562 CRISPRai Perturb-seq dataset containing 1,058 control and 1,740 perturbed cells.[14,40] Raw UMI counts were normalized to 10,000 reads per cell followed by log1p transformation.[29–32] The external dataset and model outputs were matched by case-sensitive exact gene-symbol intersection, yielding 3,015 shared features. No gene-alias conversion or missing-feature imputation was performed. Five perturbation targets were analyzed: IRF1, JUN, KLF1, LYL1, and MAP2K3, represented by 475, 362, 338, 288, and 277 perturbed cells, respectively.

For each target, Spearman correlations between the experimental response and either the additive predictor or a global-mean baseline were calculated.[6,7,14] Improvement over baseline was defined as

$$\Delta\rho_t = \rho_t^{\text{candidate}} - \rho_t^{\text{baseline}}.$$

The mean improvement across the five targets was evaluated using a one-sided exact sign-flip test. Under the null hypothesis, the signs of target-level differences were assumed exchangeable, and all $2^5 = 32$ possible sign assignments were enumerated to obtain the exact $P$ value, with a significance threshold of 0.05. Because only five targets were available, this analysis was treated as an external stress test rather than independent scientific validation.

A second external dataset was the day-8 exp8_CRISPRa_multiplexing experiment from GSE146194 (GSM4367986).[14,40] The raw matrix contained 33,862 features across 110,462 cells, including 33,694 Gene Expression and 168 CRISPR Guide Capture features. Library-size normalization was calculated using Gene Expression features only, after which the data were projected onto the fixed 5,045 Ensembl features used by the candidate model, without imputing missing features. Seven targets were analyzed—IGDCC3, IRF1, KLF1, MAP2K3, MAP2K6, MAPK1, and PRTG—comprising 1,668 perturbed and 2,517 control cells.

This dataset was used to compare the final additive candidate with the original α=1.0\alpha=1.0 additive comparator.[6,7] MSE was calculated across the fixed 5,045 features for each target and then averaged with equal weight across the seven targets. Uncertainty in the candidate–comparator MSE difference was estimated using a paired percentile bootstrap with 2,000 replicates, PCG64 random-number generation, and a fixed seed of 0; 95% percentile confidence intervals were reported. This external dataset was not used for model selection, ridge-parameter tuning, or model training. Because the experiment was measured at day 8 and did not exactly match the Norman day-5 Golden benchmark, it was treated as an external stress test rather than an independent validation dataset.

For the microbial benchmarks, condition-level Spearman and Pearson correlations were calculated across all valid targets within each growth condition, and the median condition-level Spearman correlation was used as the overall condition summary.[32] Global correlations were calculated after pooling all valid condition–target pairs, whereas target-level correlations were calculated across supported conditions for each target, requiring at least three conditions. For K562, MSE was calculated across the 5,045 output features for each perturbation condition, and macro MSE was obtained by averaging condition-level MSE values with equal weight.[6,7]

Because the microbial and K562 benchmarks represent different biological readouts and evaluation scales, all metrics were interpreted within their respective tasks and no cross-species composite score was constructed. Except for the GSE146194 bootstrap analysis, the primary benchmark results are reported as point estimates. External target-level analyses were considered descriptive or exploratory and were not subjected to additional multiple-testing correction.

**Validation evidence and reproducibility**

PRAXIS-VirtualCell classifies validation evidence into four categories: development validation, external stress testing, technical reproducibility, and scientific validation, while storing evidence state separately from software execution state.[42–46] Development validation refers to performance obtained on predefined development or Golden benchmarks; external stress testing evaluates model behavior on data from independent sources, altered experimental conditions, or shifted data distributions; technical reproducibility indicates that consistent computational results can be reproduced under fixed inputs, models, adapters, environments, and execution configurations. Scientific validation is maintained as an independent evidence state and cannot be inferred automatically from successful execution, deterministic replay, or performance on a single benchmark.

When the wild-type model is infeasible, the requested perturbation lies outside the model's declared scope, a valid feature-space mapping cannot be established, or required execution components are unavailable, the corresponding virtual experiment is recorded as unsupported or abstained and excluded from the primary performance summaries.[45,46] For requests that are technically executable but insufficiently supported by validation evidence, the system may retain the numerical prediction while distinguishing its evidence level from that of a supported scientific prediction. All real-data benchmarks and external stress tests in this study were recorded under this framework, and the scientific validation state was not automatically upgraded by any individual execution or benchmark result.

ASSOCIATED CONTENT

AUTHOR INFORMATION

**Corresponding Author**

*E-mail: xukai.jiang@sdu.edu.cn

**Author Contributions**

The manuscript was written through contributions of all authors. All authors have given approval to the final version of the manuscript.

**Notes**

The authors declare no competing financial interest.

ACKNOWLEDGMENT

This work was partly supported by the National Key Research and Development Program of China (2021YFC2103100), Shandong Excellent Young Scientists (Overseas) Fund Program (2023HWYQ-044) and Natural Science Foundation of Shandong Province (ZR2022QC014).